\documentclass[a4paper,aps,twocolumn,superscriptaddress]{revtex4}%
\usepackage{amsfonts}
\usepackage{amsmath}
\usepackage{amssymb}
\usepackage{graphicx}%
\usepackage[dvipsnames]{xcolor}
\begin{document}

\title{Phase transitions and spin-state of iron in FeO at the conditions of Earth's deep interior}

\author{E. Greenberg}\thanks{These authors contributed equally to this work.}
\affiliation{Center for Advanced Radiation Sources, University of Chicago, 5640 South Ellis Avenue, Chicago, Illinois 60637, USA}

\author{R. Nazarov}\thanks{These authors contributed equally to this work.}
\affiliation{Physics Division, Physical and Life Sciences Directorate, Lawrence Livermore National Laboratory, Livermore, CA 94551}

\author{A. Landa}
\affiliation{Physics Division, Physical and Life Sciences Directorate, Lawrence Livermore National Laboratory, Livermore, CA 94551}

\author{J. Ying}
\affiliation{Geophysical Laboratory, Carnegie Institution of Washington, Washington, DC 20015, USA}
\affiliation{Department of Physics, University of Science and Technology of China, Hefei, Anhui 230026, China}

\author{R. Q. Hood}
\affiliation{Physics Division, Physical and Life Sciences Directorate, Lawrence Livermore National Laboratory, Livermore, CA 94551}

\author{B. Hen}
\affiliation{Raymond and Beverly Sackler School of Physics and Astronomy, Tel-Aviv University, Tel Aviv 69978, Israel}

\author{R. Jeanloz}
\affiliation{Departments of Earth and Planetary Science and Astronomy, and Miller Institute for Basic Research in Science, University of California, Berkeley, California 94720, USA}

\author{V. B. Prakapenka}
\affiliation{Center for Advanced Radiation Sources, University of Chicago, 5640 South Ellis Avenue, Chicago, Illinois 60637, USA}

\author{V. V. Struzhkin}
\affiliation{Center for High Pressure Science and Technology Advanced Research, Shanghai, China}

\author{G. Kh. Rozenberg}
\affiliation{Raymond and Beverly Sackler School of Physics and Astronomy, Tel-Aviv University, Tel Aviv 69978, Israel}

\author{I. V. Leonov}
\affiliation{M.N. Miheev Institute of Metal Physics, Russian Academy of Sciences, 620108 Yekaterinburg, Russia}
\affiliation{Ural Federal University, 620002 Yekaterinburg, Russia}
\affiliation{Skolkovo Institute of Science and Technology, 143026 Moscow, Russia}

\begin{abstract}
Iron-bearing oxides undergo a series of pressure-induced electronic, spin and structural transitions that can cause seismic anomalies and dynamic instabilities in Earth's mantle and outer core. We employ x-ray diffraction and x-ray emission spectroscopy along with DFT+dynamical mean-field theory calculations to characterize the electronic structure and spin states, and crystal-structural properties of w\"ustite (Fe$_{1-x}$O) -- a basic oxide component of Earth's interior -- at high pressure-temperature conditions up to 140 GPa and 2100 K. We find that FeO exhibits complex polymorphism under pressure, with abnormal compression behavior associated with electron-spin and crystallographic phase transitions, and resulting in a substantial change of bulk modulus. Our results reveal the existence of a high-pressure phase characterized by a metallic high-spin state of iron near to the pressure-temperature conditions of Earth's core-mantle boundary. The presence of high-spin metallic iron near the base of the mantle can significantly influence the geophysical and geochemical properties of Earth's deep interior.
\end{abstract}

\maketitle


Iron monoxide, w\"ustite (Fe$_{1-x}$O), is among the most representative of compounds making up the terrestrial planets: as an electrically insulating oxide, it is akin to the constituents of rocky mantles, yet it can also be a metallic alloy, such as comprises planetary cores \cite{Ringwood77,Anderson89,Rubie04}. Because it is likely a major component of Earth's core, and may also be present near the mantle-core boundary due to interactions between Earth's rocky and liquid-metal regions, characterization of w\"ustite at high pressures and temperatures is fundamental to understanding the nature and evolution of our and other planets deep interior \cite{Yagi85,Knittle91,Fei94,Fei96,Mao96,Ozawa11b,Coppari21}. Notably, this one oxide exhibits a richness of condensed-matter phenomena that can significantly influence mantle convection, plume stability and other aspects of planetary internal dynamics, including crystal-structural phase transformations and melting, electronic (e.g., insulator-to-metal) transitions, and spin-state transitions that affect atomic structure and magnetic moments \cite{Manga96,Lin13}. An end-member of magnesiow\"ustite [(Mg, Fe)O], the second most abundant mineral of Earth's mantle,  w\"ustite exhibits (under pressure) spin-state transitions and other key features of our planet's most abundant but more complex mineral compound, (Mg,Fe)SiO$_3$ bridgmanite \cite{Lin13}.

W\"ustite has a rich phase diagram featuring at least five crystallographic phases, from the rock-salt B1 (NaCl) (or low-temperature rhombohedral rB1) to cubic B2 (CsCl) structure \cite{Yagi85,Knittle91,Fei94,Fei96,Mao96,Ozawa11b,Murakami04,Fischer11a,Fischer11b,Ohta12}. A Mott insulator with a relatively large band gap of $\sim$2.4 eV at ambient conditions, it is known to undergo a Mott insulator-to-metal phase transition (IMT) under pressure \cite{Knittle91,Ohta12,Fischer11a,Fischer11b,Leonov15,Leonov16,
Murakami04,Leonov20}. This metallization transition has been claimed to be accompanied by a high-spin (HS) to low-spin (LS) electronic transition in iron. Contraction of the iron ionic volume by up to $\sim$20-40 percent at the spin transition results in dramatic changes in seismological (density, elasticity) and transport properties (e.g., electrical and thermal conductivity), as well as in chemical partitioning of iron-bearing minerals \cite{Pasternak97,Manga96,Badro99, Lin13}. Therefore, documenting the electronic state of iron in FeO (and other iron-bearing minerals \cite{Wu_2009, Wu_2013,Hsu_2014,Wan_2022,Lavina_2016,Hu_2016,Bykova16,
Greenberg18,Koemets2021,Layek_2022}) is thought to be essential for understanding the structure and evolution of Earth's lower mantle and core-mantle boundary. However, despite being an archetype for terrestrial planetary materials, FeO is poorly understood, and its phase diagram and electronic properties remain controversial, especially at high pressure and temperatures \cite{Knittle91, Fei94, Fei96, Mao96, Murakami04, Fischer10, Ozawa11a, Ozawa11b, Fischer11a, Fischer11b, Cohen97, Ohta12, Leonov15, Leonov16,Leonov20,Pasternak97,Badro99,Lin13,Manga96,
Sun_2020,Yagi85}.


In our present study, we address these gaps in understanding and examine the phase diagram of FeO \cite{Fischer11a, Fischer11b} using state-of-the-art quantum mechanical DFT+dynamical mean-field theory (DFT+DMFT) calculations \cite{Georges96,Kotliar06,Haule07,Pourowskii07}, along with x-ray diffraction (XRD) and x-ray emission spectroscopy (XES) measurements at simultaneously high temperatures and pressures \cite{Lin05} to determine the crystal structures, electronic and spin state of FeO over a wide range of pressures.
Using DFT+DMFT it becomes possible to determine on the same footing the details of a pressure-driven Mott IMT, change in crystal structure and collapse of local moments of FeO.
%
Our results document the interplay between electron correlation and delocalization (i.e., metallic character) that -- along with changes in crystal structure and iron spin-state -- makes for rich allotropic behavior and significant variations in elastic properties.


\begin{figure}[h]
\includegraphics[width=0.45\textwidth]{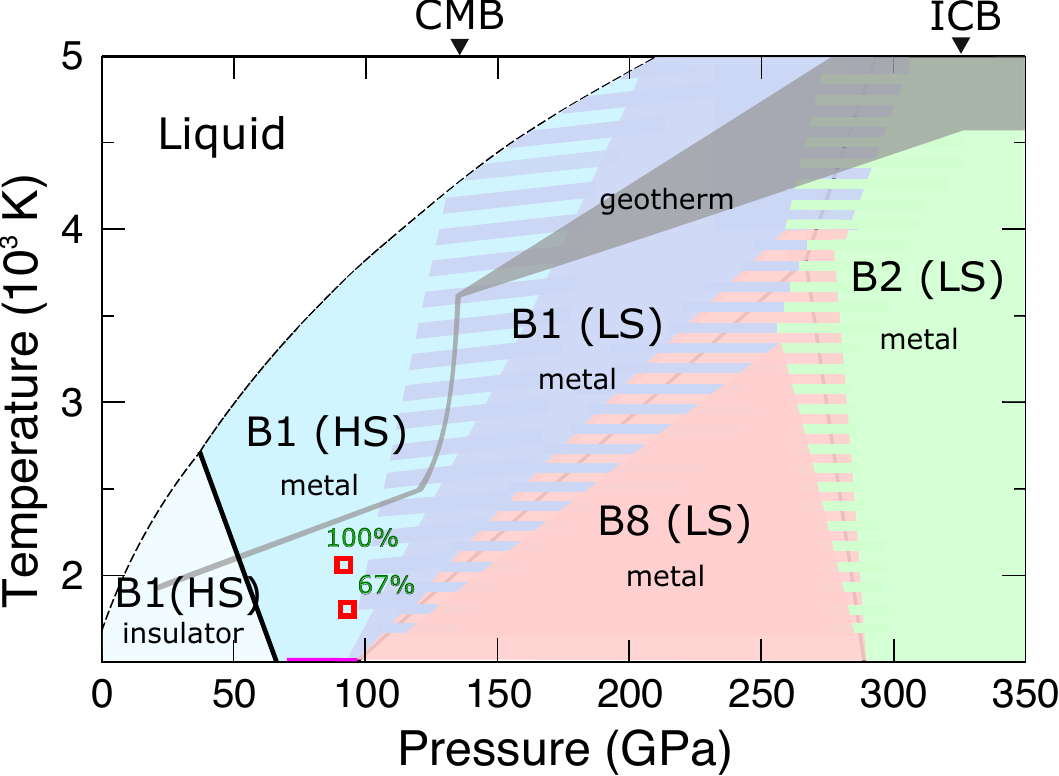}
\caption{
Pressure-temperature phase diagram of PM FeO evaluated from the DFT+DMFT crystal-structural calculations. Our results for the phase equilibria among the rock-salt B1 (NaCl), normal and inverse B8 (NiAs and anti-NiAs), and cubic B2 (CsCl) crystal phases are shown, as is the distinction between the HS and LS Fe. 
The melting curve (dashed black lines) is from previous experiments (see Ref.~\onlinecite{Fischer10} and references therein), and the pressures at the core-mantle boundary (CMB) and inner-core boundary (ICB) are indicated at the top, along with inferred temperatures inside Earth (geotherm: gray). Our XRD and XES data at pressures near 90 GPa show only the presence of the HS B1 phase at $\sim$2000 K and about 67\% HS Fe in a mixture of the rB1 and B1 structures at $\sim$1800 K (red squares).
We note that the calculated B1-B8 and B8-B2 phase boundaries are in agreement with the experimental data \cite{Fei94,Ozawa11b,Fischer11a,Ozawa_2010,Kondo_2004}.
}
\label{fig:phase_diagram}
\end{figure}

\begin{figure}[h]
\includegraphics[width=0.45\textwidth]{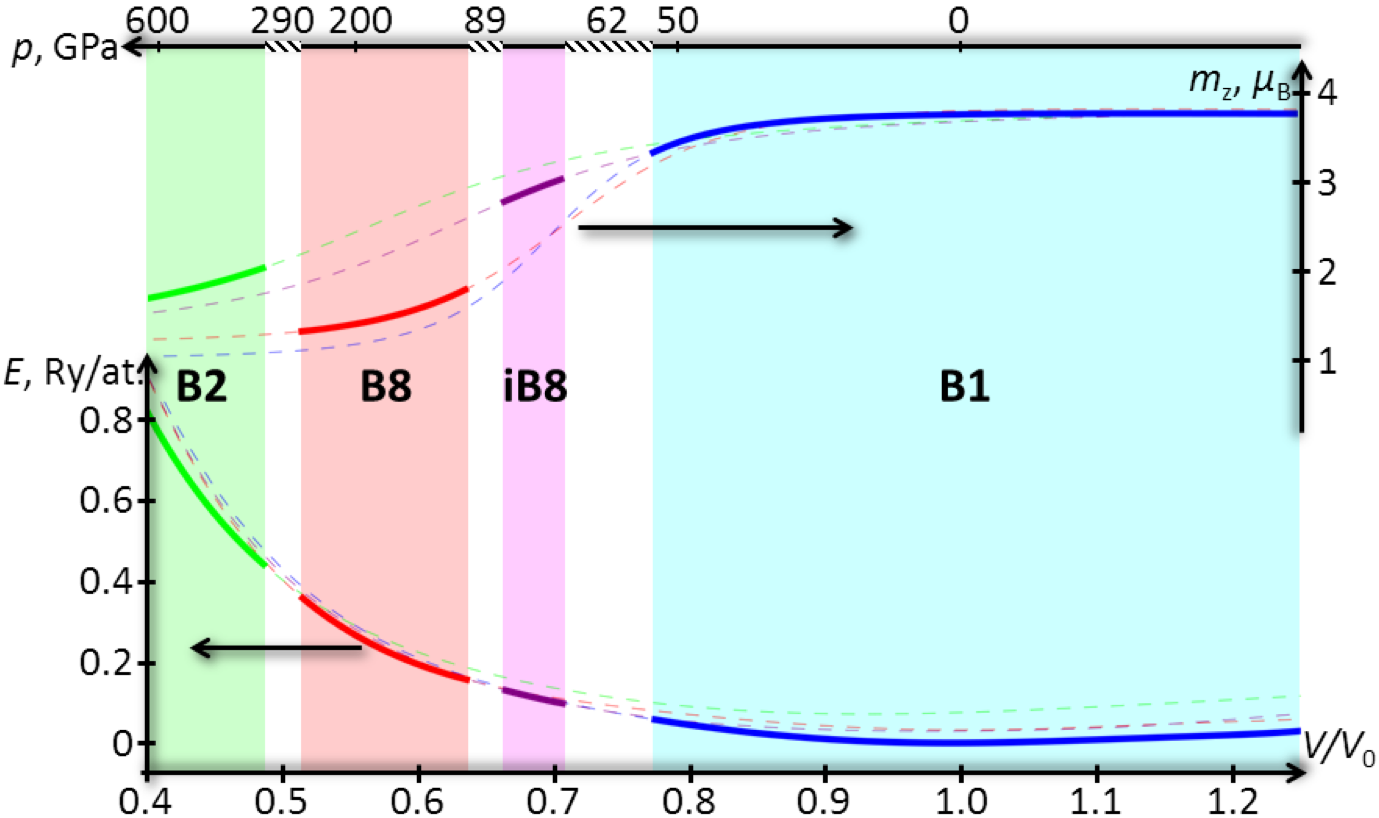}
\caption{
Total energy (bottom) and instantaneous local magnetic moment $\sqrt{\langle \hat{m}_z^2\rangle}$ (top) of PM FeO as a function of volume, as obtained by DFT+DMFT calculations for different phases at a temperature $T = 1160$ K. Colors indicate the stability ranges of different crystallographic phases.}
\label{fig:Etot}
\end{figure}

We start by computing the pressure-temperature phase diagram of paramagnetic (PM) FeO at temperatures above 1200 K using DFT+DMFT, taking into account the B1 (NaCl-type), B2 (CsCl-type), B8 (NiAs-type) and inverse B8 (Fe and O sites interchanged from B8) crystal structures (Figs.~\ref{fig:phase_diagram}, \ref{fig:Etot}). We employ a fully self-consistent in charge density DFT+DMFT method \cite{Leonov15,Leonov16,Leonov20} implemented with plane-wave pseudopotentials in DFT \cite{Baroni2001,Giannozzi2009} and continuous-time hybridization-expansion quantum Monte-Carlo algorithm in DMFT \cite{Gull2011}. In the DFT+DMFT calculations for the Fe $3d$ and O $2p$ valence states we construct a basis set of atomic-centered Wannier functions within the energy window spanned by these bands \cite{Marzari2012,Anisimov2005,Korotin08}. The effects of electron correlations in the Fe $3d$ shell are described using a Coulomb interaction $U$ ranging from 7 to 9~eV, as estimated for the different crystallographic phases (near thier stabilty range) within constrained DFT \cite{Anisimov91,Korotin08}; also, we take the Hund's exchange energy $J = 0.86$~eV. In our calculations we neglect the possible pressure-dependence of the Hubbard $U$ and Hund's $J$ values \cite{Sun_2020,Timrov_2021}, whose variation upon compression is assumed to be small. Further the details of the DFT+DMFT calculations see in Supplementary Materials (SM).

In order to account for the thermal contribution to the lattice free energy it requires to evaluate the phonon dispersion relations of PM FeO within DFT+DMFT. This is still a challenge for the systems near the Mott IMT \cite{Dai03,Floris11,Leonov2012,Leonov2014,Han18,Kocer2020,
Marcondes2020,Nakano2021}. Instead, we use a more simplified Debye-Gr\"uneisen model in which the effects of quasiharmonic contributions are taken by approximate description of the Gr\"uneisen parameter using the Dugdale-McDonald's formula \cite{Moruzzi88}. It gives a reliable estimate for the evolution of the Poisson ratio, Gr\"uneisen ratio, and Debye temperature as a function of lattice volume. (We note however that this model does not include the effects of anharmonicity and vibrational instability which seems to be relevant, e.g., near the triple point). The local magnetic moments entropy contribution is estimated as $\Delta{S_{mag}}=R ~\mathrm{ln}(M_{loc} + 1)$ (see \cite{Chuang85} and reference therein), where $M_{loc} \equiv \langle \hat{m^2_z}\rangle^{1/2}$ is the instantaneous local moment of Fe ion evaluated in DMFT. Our results for the phase diagram are summarized in Fig.~\ref{fig:phase_diagram}. Our results for the phase stability, equation of state and local magnetic moments obtained from the DFT+DMFT calculations for a temperature near 1200 K are shown in Fig.~\ref{fig:Etot}.

\begin{figure}[h]
\includegraphics[width=0.5\textwidth]{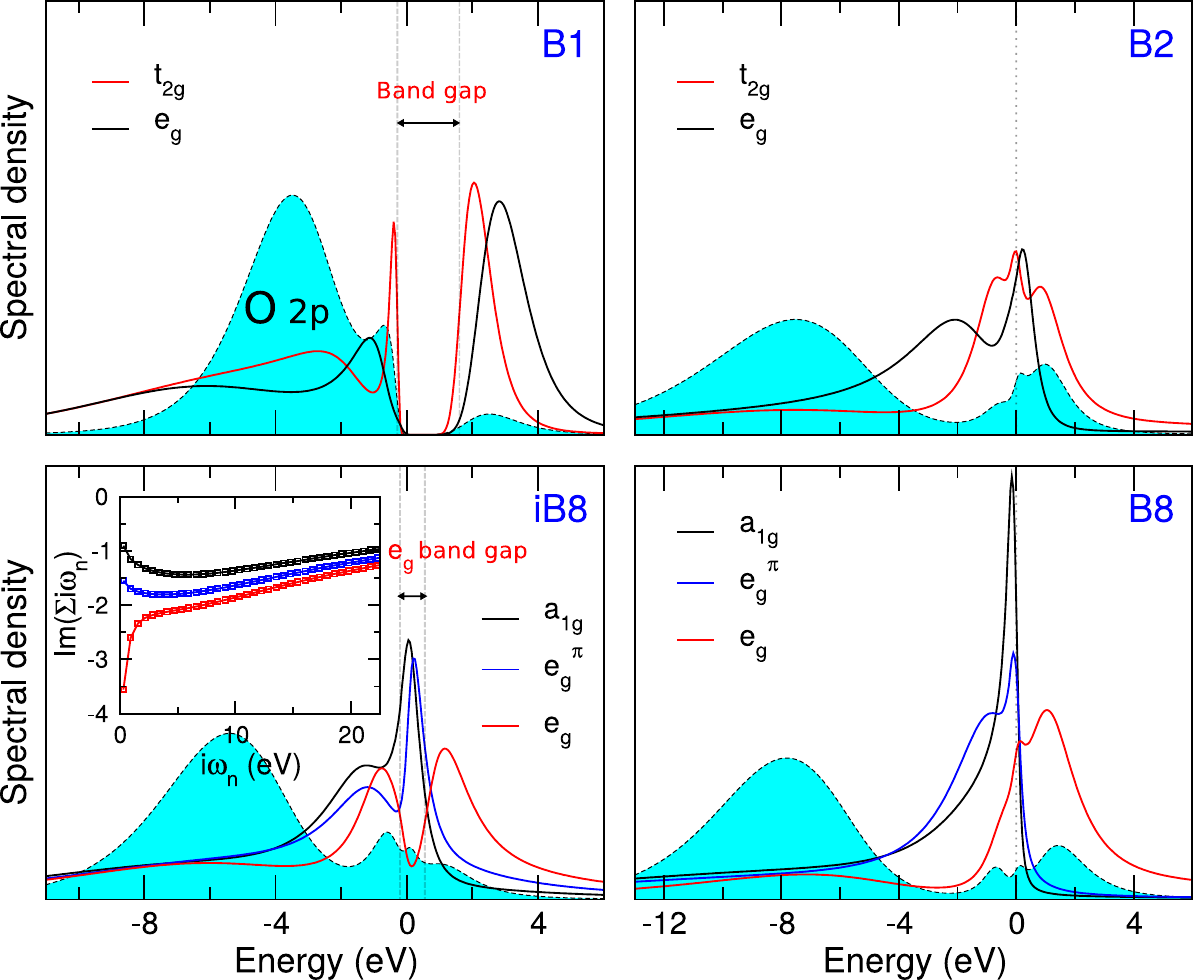}
\caption{
Partial spectral functions showing the electron-orbital contributions to bonding in different phases of PM FeO, as calculated by DFT+DMFT at $T = 1160$ K. Contributions from Fe $3d$ states ($t_{2g}$, or its split $a_{1g}$, $e_g^\pi$ components, and $e_g$ states) are shown by solid curves, and O $2p$ states are given by shaded area (cyan). 
For iB8 FeO the inset shows a Matsubara plot of the imaginary part of the self-energy, $\mathrm{Im}\Sigma(i\omega_n)$.
}
\label{fig:Spectra}
\end{figure}

Our calculations indicate that FeO is a B1-structured Mott insulator at pressures below $\sim$40-60 GPa, with a large ($\sim$2 eV) energy gap between the Fe $3d$ states (Fig.~\ref{fig:Spectra}), in agreement with previous findings \cite{Ohta12,Leonov15,Leonov16}. Fluctuating atomic-scale magnetic moments of $\sim$3.6~$\mu_B$ imply a HS $S = 2$ state for the Fe$^{2+}$ ion, as expected for the $3d^6$ configuration in an octahedral crystal field (e.g., see Fig.~2 of Ref.~\cite{Lin13}). Under pressure, FeO exhibits electronic transitions, followed by crystal-structural transformations. We find a Mott IMT for the B1 phase at about 70 GPa and 1500 K, in accord with past shock-wave and static measurements \cite{Yagi85,Knittle91,Fei94,Fei96,Mao96}. In addition, our analysis of the pressure and temperature dependence of the spectral function of FeO calculated within DFT+DMFT implies a negative Clapeyron slope for the Mott IMT, suggestive of Fermi-liquid behavior and consistent with recent experimental results \cite{Murakami04,Ozawa11a,Ozawa11b,Fischer11a,Fischer11b, Ohta12}. As the volume change is calculated to be small ($<1$\%), even a small (positive) entropy change can explain the temperature-dependence of the metallization transition.

Our results show that at high temperatures the Mott IMT does not overlap with the HS-LS transition of iron.  Instead, the spin and metallization transitions appear to be decoupled, and our calculations reveal the existence of a novel Fe HS metallic phase over the $\sim$40-150 GPa pressure range at temperatures between 1500 and 4000 K, with transition to the LS state at conditions near to those of the core-mantle boundary (Fig.~\ref{fig:phase_diagram}). 
The spin-state transition in the metallic B1 phase has a positive $P$-$T$ slope. 
While the B1 LS phase has a considerably smaller unit-cell volume and hence larger bulk modulus, by $\sim$ 9\% and 48\%, respectively, at about 110 GPa and 2500 K, we expect a broad phase boundary at the HS-LS B1 phase transition at high pressure and temperature conditions.
It is also likely that the spin transition is spread out over a finite pressure range -- hence depth interval -- in Earth's mantle, despite the sizable change of properties involved. The B1 structure remains stable at the high temperatures of Earth's mantle and outer core, and the high-spin metallic form of this structure is proposed to be stable near core-mantle boundary conditions (Fig.~\ref{fig:phase_diagram}).

Below 1500 K, the non-metallic, HS B1 phase transforms to the inverse B8 (iB8, anti-NiAs) structure above $\sim$62 GPa (Fig.~\ref{fig:Etot}). Our results suggest that the unit-cell volume drops by $\sim$10.7\% (while most likely this change spreads out over a broad pressure range), with only a slight increase in bulk modulus, from 140 to 143 GPa.  The iB8 phase is still in a HS state (local moment $\sim$3~$\mu_B$ upon compression to $\sim$0.7~$V_0$), but the trigonal prismatic coordination of the iron site causes orbital-selective collapse of the local moments \cite{Medici05}. In fact, while the Fe $a_{1g}$ states are metallic and show a quasiparticle peak at the Fermi level, a small energy gap remains for the Fe $e_g$ states (see Fig.~\ref{fig:Spectra}). It is seen as a divergence of the imaginary part of the Fe $e_g$ self-energy $\mathrm{Im}\Sigma(\omega_n)$ at the lowest Matsubara frequencies (see the inset of Fig.~\ref{fig:Spectra}).

\begin{table}[h]
\begin{ruledtabular}
\begin{tabular}{lcccccc}
  & $V_0^\mathrm{HS}$  & $K_{0,T}^\mathrm{HS}$ & $V_0^\mathrm{LS}$  & $K_{0,T}^\mathrm{LS}$ & $P_\mathrm{tr.}^\mathrm{HS-LS}$ & $\Delta V/V$ (\%) \\
\hline
B1	 & 144.1	& 140	& 122.4	& 210 &	73 &	9 \\
B2	 & 133.7 &	136	& 110.8 &	256	& 116 &	6 \\
iB8 & 138.1 &	143	& 119.6	& 227 &	62 &	7 \\
B8	 & 143.3	& 128	& 114	& 274	& 43 &	13 \\
\end{tabular}
\caption{\label{tab:tab1}
 Parameters of the third-order Birch-Murnaghan equation of states of PM FeO phases, as evaluated from the DFT+DMFT total energy results at an electronic temperature $T = 1160$ K. $V_0$ is volume and $K_{0,T}$ isothermal bulk modulus (subscript zero indicates zero pressure, $P = 0$, and $dK_{0,T}/dP$ is fixed to 4.1).}
\end{ruledtabular}
\end{table}

The iB8 phase then transforms to the normal B8 (NiAs) structure above 89 GPa (at $\sim$1200~K), with a small reduction in unit-cell volume ($\sim$2.1\%) but a near doubling of the bulk modulus to 274 GPa. The transition involves a collapse of local magnetic moments into the LS state ($\sim$0.9 $\mu_B$, see Fig.~\ref{fig:Etot}), and appearance of correlated metallic behavior. The B2 (CsCl) structure, which is stable above 290 GPa at $\sim$1200 K, is also a LS, correlated-electron metal, as indicated by the lack of a band gap in the spectral density and the low value of local moments (Figs.~\ref{fig:Etot}, \ref{fig:Spectra}). 

Our results thus suggest that w\"ustite, which may appear in the lowermost mantle due to chemical reactions at the core-mantle boundary, is metallic and could produce seismological anomalies caused by spin-state transitions at deep-mantle depths ($\sim$1700-2000 km). Enhanced electrical conductivity due to the presence of metallic FeO would influence heat transfer from Earth's core into the mantle, as well as the temporal evolution of magnetic field lines crossing into the lower mantle. From the perspective of phase stability, it is notable that the metallization and spin-state transitions are \emph{decoupled} for the B1 phase of FeO at high temperatures (Fig.~\ref{fig:phase_diagram}).


\begin{figure}[h]
\includegraphics[width=0.45\textwidth]{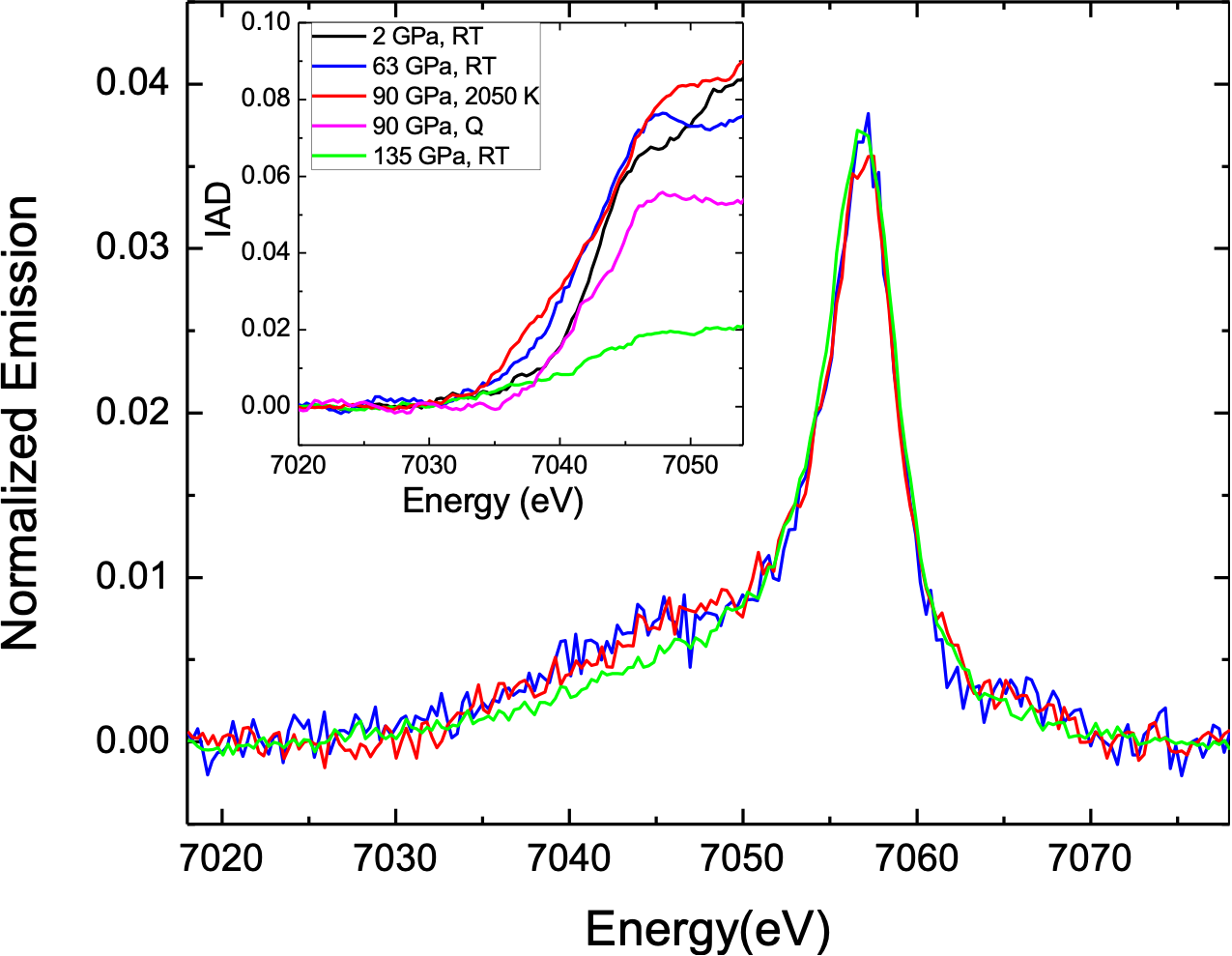}
\caption{
Fe $K_\beta$ x-ray emission spectra collected from FeO at high pressure as a function of temperature. The spectra have been normalized and shifted so that the main peak is centered around 7058 eV. Inset: Integrated absolute difference (IAD) for a few of the collected spectra. The abundance of the HS state remains close to 100\% up to 74 GPa, and decreases appreciably upon further compression [90 GPa quenched (Q) and 135 GPa room temperature (RT)]. We observe complete recovery (with 100\% abundance) of the HS state, and completion of the rB1-B1 structural transition, upon heating to 2050 K (spectrum at 90 GPa, 2050 K).
}
\label{fig:XES}
\end{figure}

To verify these results, and specifically to confirm the existence of the HS Fe in the metallic B1 phase, we used XES and XRD to characterize FeO at 60-140 GPa and temperatures up to 2100 K (Fig.~\ref{fig:XES}). Such a combined experimental study is technically challenging, and has rarely been attempted \cite{Lin05,Lin07}.  Our x-ray spectroscopy was conducted on samples with stoichiometry Fe$_{0.94}$O (see SM for the details) at beamline 13-IDD of the Advanced Photon Source, Argonne National Laboratory. For combined XRD and XES measurements at high pressures and temperatures, a XES system was added to the 13-IDD beamline that includes laser heating for XRD with diamond-anvil cells. Our diffraction measurements were performed at a wavelength of $\lambda = 0.4959$ \AA\ with a Mar165 CCD detector. 
Iron $K_{\beta 1,3}$ spectra were collected using an excitation energy of 10.75~keV with a spot size of $\sim 4 \times 4$ $\mu$m, and a curved Si 440 analyzer ($a = 5.431$ \AA) in a Rowland circle spectrometer geometry having nominal spherical diameter of $\sim$1000 mm (an Fe wire is used to calibrate the analyzer angle of 66.18$^{\circ}$, assuming that the main peak is at 7058~eV). Spectra were collected from 7018 to 7078~eV in $\sim$0.3 eV steps (1-10 s collection time for each step), with each measurement repeated an even number of times (2-10). At each step the entire region-of-interest on the detector is summed, and the final spectrum is a summation of all repetitions. A sample of siderite FeCO$_3$ at 60 GPa was used as a low-spin standard \cite{Mattila07a} in order to evaluate the integrated absolute difference (IAD) between the sample and the LS FeCO$_3$ as $\mathrm{IAD}=\int_{E_i}^{E_f}~|I^\mathrm{shifted}_{norm}(E)-I^\mathrm{FeCO_3}_{norm}(E)|dE$, where $I_{norm}(E)$ is the normalized to a unit background-subtracted spectrum $I_{norm}(E)=I(E)/\int^{7065}_{7018}|I(E)|dE$; $E_i\sim7030$~eV is just before the satellite peak, $E_f\sim 7045$~eV, above (see SM).

\begin{figure}[h]
\includegraphics[width=0.45\textwidth]{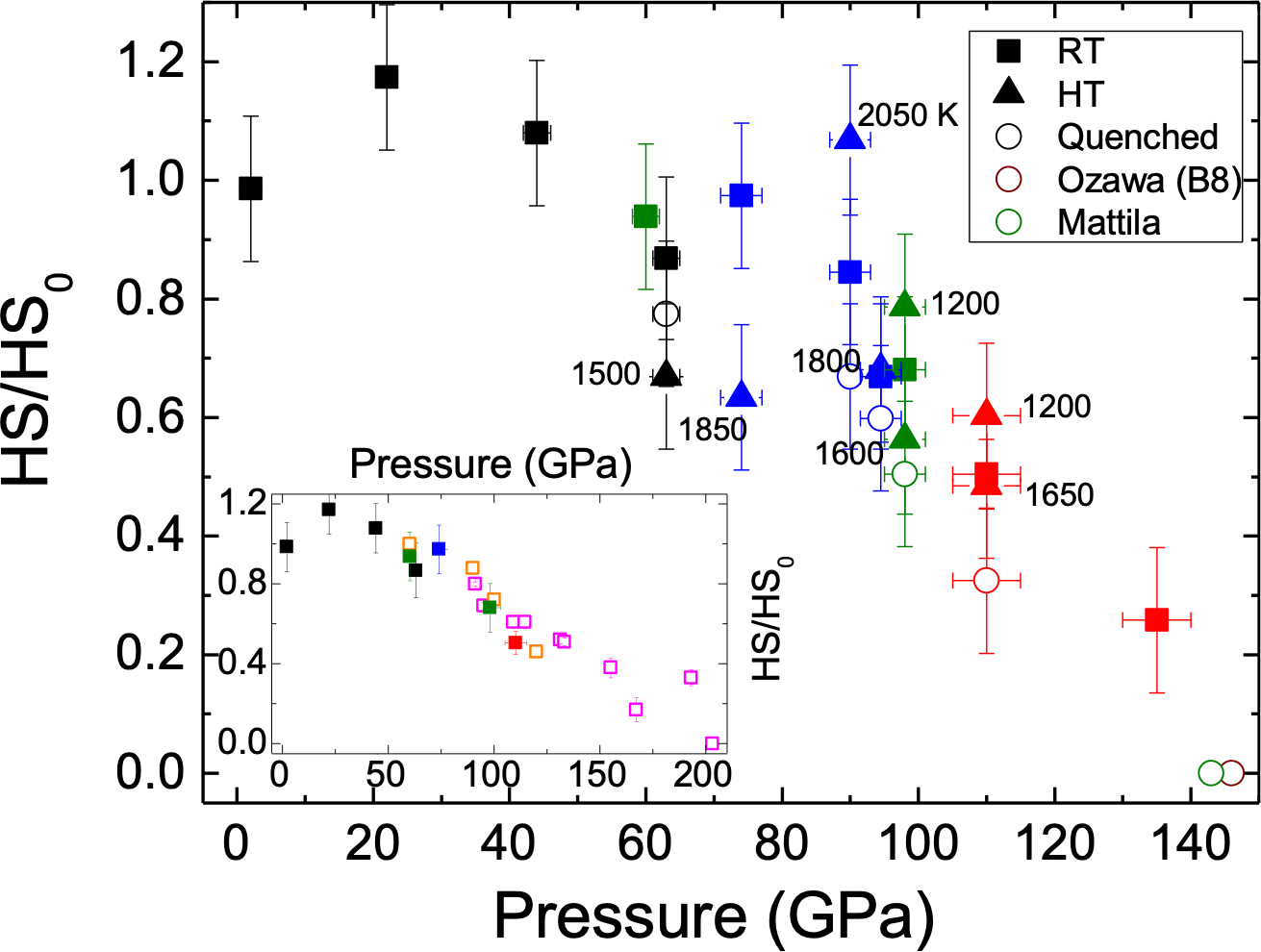}
\caption{
A relative abundance of the Fe HS state to that at ambient pressure (HS$_0$) at high pressures and either room temperature (RT) or high temperature (HT), as calculated from IAD obtained from XES (squares, triangles and circles represent RT, HT and quenched data, respectively). Colors represent data collected from different sample runs. Note that Ozawa \emph{et al.} and Mattila \emph{et al.'s} data were obtained after laser-heating, resulting in a structural transition into the B8 phase \cite{Ozawa11a,Mattila07b}. Inset: HS/HS$_0$ abundance at RT (unheated samples), revealing the sluggish spin transition induced by pressure alone. Open magenta and orange squares represent the RT M\"ossbauer spectroscopy results obtained by Hamada \emph{et al.} \cite{Hamada16} and Pasternak \emph{et al.} \cite{Pasternak97}.}
\label{fig:Abundance}
\end{figure}

At room temperature, we find the iron to be in the HS state up to 74 GPa, with the LS Fe appearing on further compression: the ratio of high- to low-spin iron in FeO drops to about 50\% by 110 GPa (Fig.~\ref{fig:Abundance}). This is in agreement with M\"ossbauer measurements \cite{Pasternak97} and recent DFT+DMFT calculations \cite{Ohta12,Leonov15,Leonov16,Leonov20}, but is not entirely consistent with prior XES work \cite{Badro99}. We attribute this difference to the previous x-ray spectroscopy having probed the sample from the side (through a Be gasket), across the sample's full pressure distribution, making it difficult to see the disappearance of the $K_\beta$ shoulder that is the signature of the LS state. We note that FeO is in the distorted (rB1) form of the B1 structure over most of this pressure range \cite{Yagi85, Fischer11a, Fischer11b}, with the sluggish spin transition starting in the rB1 phase and being followed by transformation to the B8 phase at room temperature. Therefore, consistent with our DFT+DMFT calculations we conclude that the electronic-spin and structural transitions are decoupled.

With increasing temperature near 90 GPa, XRD shows the presence of the low-temperature rB1 structure up to about 1800~K, and completion of the rB1-to-B1 phase transition -- with the iron 100\% in the HS state -- upon heating to 2050 K (Fig.~\ref{fig:phase_diagram}). Our experiments do not confirm the previously reported LS B1 phase of FeO above $\sim$70 GPa at temperatures above 1300~K \cite{Fischer11b,Ohta12,Cohen97,Leonov15}, but they are in qualitative agreement
with our DFT+DMFT calculations.  Furthermore, we note the sensitivity of the spin state to changes in the crystal-field splitting of the Fe $3d$ energy levels across the rB1-B1 phase transformation of FeO. 


We thus conclude from both experiment and theory that a novel HS B1-type phase of FeO appears near conditions relevant to Earth's core-mantle boundary. Therefore, if present in this region, as might be expected due to chemical reactions between mantle rock and liquid metal of the core, FeO can contribute to the seismological anomalies of the D'' region, and cause lateral variations in heat flow into the base of the mantle (e.g., Ref.~\cite{Manga96}).

Our DFT+DMFT results are for crystalline phases, so it is premature to make quantitative predictions about the properties of Earth's fluid outer core \cite{Morard_2022}. Nevertheless, the evidence we have found for significant changes in bulk sound velocity  ($\sqrt{K}/\rho$) across both structural (B8-B2: about -5\%) and electronic (HS-LS: about +10\%) transitions in metallic FeO raises the possibility of anomalous variations in seismic-wave velocity with depth through the core. Specifically, it may be imprudent to treat the outer core as though it followed the isentropic equation of state of a single phase, because changes in liquid and electronic structures could cause subtle variations in the depth dependence of seismic-wave velocity throughout this region. Our results also prompt further investigation of the electronic states of Earth's lower-mantle minerals as possible sources of seismic and chemical heterogeneity, as well as thermal instabilities.

\begin{acknowledgments}
The authors would like to thank H. Yang, J.F. Lin and J. Liu for their assistance in preparations for the experiment, C. Kenney-Benson for assistance with handling Be, and
P. Chow for assistance in preparing the XES setup at the beamline. 
Portions of this work were performed at GeoSoilEnviroCARS (The University of Chicago, Sector 13), Advanced Photon Source (APS), Argonne National Laboratory.  GeoSoilEnviroCARS is supported by the National Science Foundation-Earth Sciences (EAR-1634415) and Department of Energy-GeoSciences (DE-FG02-94ER14466). This research used resources of the Advanced Photon Source, a U.S. Department of Energy (DOE) Office of Science User Facility operated for the DOE Office of Science by Argonne National Laboratory under Contract No. DE-AC02-06CH11357. A portion of this work was performed at HPCAT (Sector 16), Advanced Photon Source (APS), Argonne National Laboratory. HPCAT operations are supported   by DOE-NNSA under Award No. DE-NA0001974, with partial instrumentation funding by NSF. The Advanced Photon Source is a U.S. Department of Energy (DOE) Office of Science User Facility operated for the DOE Office of Science by Argonne National Laboratory under Contract No. DE-AC02-06CH11357.
Computing support for this work (R.N., A.L., and R.Q.H.) came from the LLNL Computing Grand Challenge program. This work performed under the auspices of the U.S. DOE by LLNL under Contract DE-AC52-07NA27344.
R.J. acknowledges support from the U.S. Department of Energy and University of California. 
This research was supported in part by Israeli Science Foundation Grant 
\#1189/14, \#1552/18, and \#1748/20. I.V.L. acknowledges support by the state assignment of Minobrnauki of Russia (Theme ``Electron'' No. 122021000039-4). Theoretical analysis of structural properties was supported by Russian Science Foundation (project No. 19-72-30043).
\end{acknowledgments}





\end{document}